\documentclass[12pt]{article}
\usepackage{amssymb}
\usepackage{amsfonts}
\usepackage{amsmath}
\usepackage{eurosym}
\usepackage{makeidx}
\usepackage{amssymb}
\usepackage{times}
\usepackage{anysize}
\usepackage{cite}
\usepackage{lineno}

\input{tcilatex}
\begin{document}

\title{Macroscopic Dynamics of the Strong-Coupling BCS-Hubbard Model}
\author{J.-B. Bru \and W. de Siqueira Pedra}
\maketitle

\begin{abstract}
The aim of the current paper is to illustrate, in a simple example, our
recent, very general, rigorous results \cite{BruPedra-MFII,BruPedra-MFIII}
on the dynamical properties of fermions and quantum-spin systems with
long-range, or mean-field, interactions, in infinite volume. We consider
here the strong-coupling BCS-Hubbard model studied in \cite%
{BruPedra1,Bru-pedra-MF-IV}, because this example is very pedagogical and,
at the same time, physically relevant for it highlights the impact of the
(screened) Coulomb repulsion on ($s$-wave) superconductivity. \bigskip

\end{abstract}

\section{Presentation of the Model\label{Presentation of the Model}}

The most general form of a translation invariant model for fermions with
two-body interactions\ in a cubic box $\Lambda _{L}\doteq \{\mathbb{Z}\cap %
\left[ -L,L\right] \}^{d}$ ($d$-dimensional crystal) of volume $|\Lambda
_{L}|$, $L\in \mathbb{N}_{0}$, is given in momentum space by%
\begin{equation}
\mathrm{H}_{L}^{Full}=\underset{k\in \Lambda _{L}^{\ast },\ \mathrm{s}\in 
\mathrm{S}}{\sum }\left( \varepsilon _{k}-\mu \right) \tilde{a}_{k}^{\ast }%
\tilde{a}_{k}+\frac{1}{\left\vert \Lambda _{L}\right\vert }\underset{\mathrm{%
s}_{1},\mathrm{s}_{2},\mathrm{s}_{3},\mathrm{s}_{4}\in \mathrm{S}}{\underset{%
k,k^{\prime },q\in \Lambda _{L}^{\ast }}{\sum }}g_{\mathrm{s}_{1},\mathrm{s}%
_{2},\mathrm{s}_{3},\mathrm{s}_{4}}\left( k,k^{\prime },q\right) \tilde{a}%
_{k+q,\mathrm{s}_{1}}^{\ast }\tilde{a}_{k^{\prime }-q,\mathrm{s}_{2}}^{\ast }%
\tilde{a}_{k^{\prime },\mathrm{s}_{3}}\tilde{a}_{k,\mathrm{s}_{4}}\ .
\label{hamil general}
\end{equation}%
See \cite[Eq. (2.1)]{Metzner}. Here, $\mathrm{S}$ is some finite (spin) set
representing the internal degrees of freedom of quantum particles and $%
\Lambda _{L}^{\ast }$ is the reciprocal lattice of quasi-momenta (periodic
boundary conditions) associated with $\Lambda _{L}$. The operator $\tilde{a}%
_{k,\mathrm{s}}^{\ast }$ (respectively $\tilde{a}_{k,\mathrm{s}}$) creates
(respectively annihilates) a fermion with spin $\mathrm{s}\in \mathrm{S}$
and (quasi-) momentum $k\in \Lambda _{L}^{\ast }$, the function $\varepsilon
_{k}$ represents the kinetic energy of a fermion with (quasi-) momentum $k$
and the real number $\mu $ is the chemical potential. The last term of (\ref%
{hamil general}) corresponds to a translation-invariant two-body interaction
written in the momentum space.

One important example of a fermionic system with long-range interactions is
given in the scope of the celebrated BCS theory -- proposed in the late
1950s (1957) to explain conventional type I superconductors. The lattice
version of this theory is obtained from (\ref{hamil general}) by taking $%
\mathrm{S}\doteq \{\uparrow ,\downarrow \}$ and imposing 
\begin{equation*}
g_{\mathrm{s}_{1},\mathrm{s}_{2},\mathrm{s}_{3},\mathrm{s}_{4}}\left(
k,k^{\prime },q\right) =\delta _{k,-k^{\prime }}\delta _{\mathrm{s}%
_{1},\uparrow }\delta _{\mathrm{s}_{2},\downarrow }\delta _{\mathrm{s}%
_{3},\downarrow }\delta _{\mathrm{s}_{4},\uparrow }f\left( k,-k,q\right) 
\end{equation*}%
for some function $f$: It corresponds to the so-called (reduced) BCS\
Hamiltonian%
\begin{equation}
\mathrm{H}_{L}^{BCS}\doteq \sum\limits_{k\in \Lambda _{L}^{\ast }}\left(
\varepsilon _{k}-\mu \right) \left( \tilde{a}_{k,\uparrow }^{\ast }\tilde{a}%
_{k,\uparrow }+\tilde{a}_{k,\downarrow }^{\ast }\tilde{a}_{k,\downarrow
}\right) -\frac{1}{\left\vert \Lambda _{L}\right\vert }\sum_{k,q\in \Lambda
_{L}^{\ast }}\gamma _{k,q}\tilde{a}_{k,\uparrow }^{\ast }\tilde{a}%
_{-k,\downarrow }^{\ast }\tilde{a}_{-q,\downarrow }\tilde{a}_{q,\uparrow }\ ,
\label{BCS Hamilt}
\end{equation}%
where $\gamma _{k,q}$ is a positive\footnote{%
The positivity of $\gamma _{k,q}$ imposes constraints on the choice of the
function $f$.} function. Because of the term $\delta _{k,-k^{\prime }}$, the
interaction of this model has a long-range character, in position space. The
simple choice $\gamma _{k,q}=\gamma >0$ in (\ref{BCS Hamilt}) is still
physically very interesting since, even when $\varepsilon _{k}=0$, the BCS\
Hamiltonian qualitatively displays most of basic properties of real
conventional type I superconductors. See, e.g. \cite[Chapter VII, Section 4]%
{Thou}. The case $\varepsilon _{k}=0$ is known as the strong coupling limit
of the BCS model. The dynamical properties of the BCS\ Hamiltonian $\mathrm{H%
}_{L}^{BCS}$ with $\gamma _{k,q}=\gamma >0$ can be \emph{explicitly}
computed from results of \cite{BruPedra-MFII,BruPedra-MFIII}, but we prefer
here to consider another BCS-type model including the Hubbard interaction,
this being a much richer example.

An important physical fact not taken into account in the BCS theory is the
Coulomb interaction between electrons or holes, which can imply strong
correlations, like in cuprates with the universally observed Mott transition
at zero doping. This problem was of course already addressed in theoretical
physics right after the emergence of the Fr\"{o}hlich model and the BCS
theory, see, e.g., \cite{Bogoliubov-tolman shirkov}.

We present below a model, named here the strong-coupling BCS-Hubbard
Hamiltonian, which is rigorously studied at equilibrium in \cite{BruPedra1}
in order to understand the possible thermodynamic impact of the Coulomb
repulsion on ($s$-wave) superconductivity. An interesting mathematical
outcome of \cite{BruPedra1} on the strong-coupling BCS-Hubbard Hamiltonian
is the existence of a superconductor-Mott insulator phase transition, like
in cuprates which must be doped to become superconductors.

The results of \cite{BruPedra1} are based on an \emph{exact} study of the
phase diagram of the strong-coupling BCS-Hubbard model defined, in a cubic
box $\Lambda _{L}\doteq \{\mathbb{Z}\cap \left[ -L,L\right] \}^{d}$ ($d\in 
\mathbb{N}$) of volume $|\Lambda _{L}|$ for $L\in \mathbb{N}_{0}$, by the
Hamiltonian%
\begin{equation}
\mathrm{H}_{L}\doteq \sum_{x\in \Lambda _{L}}\left( 2\lambda n_{x,\uparrow
}n_{x,\downarrow }-\mu \left( n_{x,\uparrow }+n_{x,\downarrow }\right)
-h\left( n_{x,\uparrow }-n_{x,\downarrow }\right) \right) -\frac{\gamma }{%
\left\vert \Lambda _{L}\right\vert }\sum_{x,y\in \Lambda _{L}}a_{x,\uparrow
}^{\ast }a_{x,\downarrow }^{\ast }a_{y,\downarrow }a_{y,\uparrow }
\label{strong coupling ham}
\end{equation}%
for real parameters $\mu ,h\in \mathbb{R}$ and $\lambda ,\gamma \geq 0$. The
operator $a_{x,\mathrm{s}}^{\ast }$ (resp. $a_{x,\mathrm{s}}$) creates
(resp. annihilates) a fermion with spin $\mathrm{s}\in \{\uparrow
,\downarrow \}$ at lattice position $x\in \mathbb{Z}^{d}$, $d=1,2,3,...,$
whereas $n_{x,\mathrm{s}}\doteq a_{x,\mathrm{s}}^{\ast }a_{x,\mathrm{s}}$ is
the particle number operator at position $x$ and spin $\mathrm{s}$. They are
linear operators acting on the fermion Fock space $\mathcal{F}_{\Lambda
_{L}} $, where%
\begin{equation}
\mathcal{F}_{\Lambda }\doteq \bigwedge \mathbb{C}^{\Lambda \times \{\uparrow
,\downarrow \}}\equiv \mathbb{C}^{2^{\Lambda \times \{\uparrow ,\downarrow
\}}}  \label{fock}
\end{equation}%
for any $\Lambda \subseteq \mathbb{Z}^{d}$ and $d\in \mathbb{N}$. The first
term of the right-hand side of (\ref{strong coupling ham}) represents the
(screened) Coulomb repulsion as in the celebrated Hubbard model. The second
term corresponds to the strong-coupling limit of the kinetic energy, also
called \textquotedblleft atomic limit\textquotedblright\ in the context of
the Hubbard model. The third term is the interaction between spins and the
external magnetic field $h$. The last term is the BCS interaction written in
the $x$-space since%
\begin{equation}
\frac{\gamma }{\left\vert \Lambda _{L}\right\vert }\sum_{x,y\in \Lambda
_{L}}a_{x,\uparrow }^{\ast }a_{x,\downarrow }^{\ast }a_{y,\downarrow
}a_{y,\uparrow }=\frac{\gamma }{\left\vert \Lambda _{L}\right\vert }%
\sum_{k,q\in \Lambda _{L}^{\ast }}\tilde{a}_{k,\uparrow }^{\ast }\tilde{a}%
_{-k,\downarrow }^{\ast }\tilde{a}_{q,\downarrow }\tilde{a}_{-q,\uparrow }\ .
\label{BCS interactions}
\end{equation}%
See (\ref{BCS Hamilt}) with $\gamma _{k,q}=\gamma >0$. This homogeneous BCS
interaction should be seen as a long-range effective interaction, the
precise mediators of which are not relevant, i.e., they could be phonons, as
in conventional type I superconductors, or anything else.

\section{Approximating Hamiltonians}

The thermodynamic impact of the Coulomb repulsion on s-wave superconductors
is analyzed in \cite{BruPedra1}, via a rigorous study of equilibrium and
ground states of the strong-coupling BCS-Hubbard Hamiltonian: An Hamiltonian
like $\mathrm{H}_{L}$ defines in the thermodynamic limit $L\rightarrow
\infty $ a free-energy density functional on a suitable set of states of the
CAR algebra of the lattice $\mathbb{Z}^{d}$. See \cite[Section 6.2]%
{BruPedra1} for more details. Minimizers $\omega $ of the free-energy
density are called equilibrium states of the model and, for any $L\in 
\mathbb{N}_{0}$, the Gibbs states $\omega ^{(L)}$, defined on the algebra $%
\mathcal{B}(\mathcal{F}_{\Lambda _{L}})$ of linear operators acting on the
fermion Fock space $\mathcal{F}_{\Lambda _{L}}$ (\ref{fock}) by 
\begin{equation}
\omega ^{(L)}\left( A\right) \doteq \mathrm{Trace}_{\mathcal{F}_{\Lambda
_{L}}}\left( A\frac{\mathrm{e}^{-\beta \mathrm{H}_{L}}}{\mathrm{Trace}_{%
\mathcal{F}_{\Lambda _{L}}}\left( \mathrm{e}^{-\beta \mathrm{H}_{L}}\right) }%
\right) \ ,\qquad A\in \mathcal{B}\left( \mathcal{F}_{\Lambda _{L}}\right) \
,  \label{gibbs1}
\end{equation}%
at inverse temperature $\beta >0$, converges\footnote{%
In the weak$^{\ast }$ topology.} in the thermodynamic limit $L\rightarrow
\infty $ to a well-defined equilibrium state. The important point in such an
analysis is the study of a variational problem over complex numbers: By the
so-called approximating Hamiltonian method \cite%
{approx-hamil-method0,approx-hamil-method,approx-hamil-method2} one uses an
approximation of the Hamiltonian, which, in the case of the strong-coupling
BCS-Hubbard Hamiltonian, is equal to the $c$-dependent Hamiltonian%
\begin{equation}
\mathrm{H}_{L}\left( c\right) \doteq \sum_{x\in \Lambda _{L}}\left( 2\lambda
n_{x,\uparrow }n_{x,\downarrow }-\mu \left( n_{x,\uparrow }+n_{x,\downarrow
}\right) -h\left( n_{x,\uparrow }-n_{x,\downarrow }\right) -\gamma \left(
ca_{x,\uparrow }^{\ast }a_{x,\downarrow }^{\ast }+\bar{c}a_{x,\downarrow
}a_{x,\uparrow }\right) \right)   \label{Hamiltonian BCS-Hubbard approx}
\end{equation}%
with $c\in \mathbb{C}$. The main advantage of using this $c$-dependent
Hamiltonian, in comparison with $\mathrm{H}_{L}$, is the fact that it is a
sum of shifts of the same on-site operator. For an appropriate choice of
(order) parameter $c\in \mathbb{C}$, it leads to the exact thermodynamics of
the strong-coupling BCS-Hubbard model, in the limit $L\rightarrow \infty $:
At inverse temperature $\beta >0$,%
\begin{equation}
\lim_{L\rightarrow \infty }\frac{1}{\beta \left\vert \Lambda _{L}\right\vert 
}\ln \mathrm{Trace}_{\mathcal{F}_{\Lambda _{L}}}\left( \mathrm{e}^{-\beta 
\mathrm{H}_{L}}\right) =\underset{c\in \mathbb{C}}{\sup }\left\{ -\gamma
|c|^{2}+\lim_{L\rightarrow \infty }\left\{ \frac{1}{\beta \left\vert \Lambda
_{L}\right\vert }\ln \mathrm{Trace}_{\mathcal{F}_{\Lambda _{L}}}\left( 
\mathrm{e}^{-\beta \mathrm{H}_{L}\left( c\right) }\right) \right\} \right\} 
\label{var pb}
\end{equation}%
and the (exact) Gibbs state $\omega ^{(L)}$ converges\footnote{%
In the weak$^{\ast }$ topology.} to a convex combination\footnote{%
More precisely, it converges to the barycenter of a Choquet measure.} of the
thermodynamic limit $L\rightarrow \infty $ of the (approximating) Gibbs
state $\omega ^{(L,\mathfrak{d})}$ defined by 
\begin{equation}
\omega ^{(L,\mathfrak{d})}\left( A\right) \doteq \mathrm{Trace}_{\mathcal{F}%
_{\Lambda _{L}}}\left( A\frac{\mathrm{e}^{-\beta \mathrm{H}_{L}\left( 
\mathfrak{d}\right) }}{\mathrm{Trace}_{\mathcal{F}_{\Lambda _{L}}}\left( 
\mathrm{e}^{-\beta \mathrm{H}_{L}\left( \mathfrak{d}\right) }\right) }%
\right) \ ,\qquad A\in \mathcal{B}\left( \mathcal{F}_{\Lambda _{L}}\right) \
,  \label{gibbs2}
\end{equation}%
the complex number $\mathfrak{d}\in \mathbb{C}$ being a solution to the
variational problem (\ref{var pb}). Since $\gamma \geq 0$, this can be
heuristically be seen from the inequality 
\begin{equation*}
\gamma \left\vert \Lambda _{L}\right\vert \left\vert c\right\vert ^{2}+%
\mathrm{H}_{L}\left( c\right) -\mathrm{H}_{L}=\gamma \left( \mathfrak{c}%
_{0}^{\ast }-\sqrt{\left\vert \Lambda _{L}\right\vert }\bar{c}\right) \left( 
\mathfrak{c}_{0}-\sqrt{\left\vert \Lambda _{L}\right\vert }c\right) \geq 0\ ,
\end{equation*}%
where 
\begin{equation}
\mathfrak{c}_{0}\doteq \frac{1}{\sqrt{\left\vert \Lambda _{L}\right\vert }}%
\sum_{x\in \Lambda _{L}}a_{x,\downarrow }a_{x,\uparrow }
\label{dynamics approx00}
\end{equation}%
(resp. $\mathfrak{c}_{0}^{\ast }$) annihilates (resp. creates) one Cooper
pair within the condensate, i.e., in the zero-mode for fermion pairs. This
suggests the proven fact \cite[Theorem 3.1]{BruPedra1} that 
\begin{equation}
\left\vert \mathfrak{d}\right\vert ^{2}=\lim_{L\rightarrow \infty }\frac{%
\omega ^{(L)}\left( \mathfrak{c}_{0}^{\ast }\mathfrak{c}_{0}\right) }{%
\left\vert \Lambda _{L}\right\vert }  \label{dynamics approx0}
\end{equation}%
for any\footnote{%
This implies that any solution $\left\vert d\right\vert $ to the variational
problem (\ref{var pb}) must have the same absolute value.} $\mathfrak{d}\in 
\mathbb{C}$ solution to the variational problem (\ref{var pb}). The
parameter [**$\left\vert \mathfrak{d}\right\vert ^{2}$**] is the condensate
density of Cooper pairs and so,[** $\mathfrak{d}\neq 0$**] corresponds to
the existence of a superconducting phase, which is shown to exist for
sufficiently large $\gamma \geq 0$. See also \cite[Figs. 1,2,3]{BruPedra1}.

\section{Dynamical Problem in the Thermodynamic Limit}

As is usual, a Hamiltonian like the strong-coupling BCS-Hubbard model drives
a dynamics in the Heisenberg picture of quantum mechanics: The corresponding
time-evolution is, for $L\in \mathbb{N}_{0}$, a continuous group $\{\tau
_{t}^{(L)}\}_{t\in {\mathbb{R}}}$ of $\ast $-automorphisms of the algebra $%
\mathcal{B}(\mathcal{F}_{\Lambda _{L}})$ of linear operators acting on the
Fermion Fock space $\mathcal{F}_{\Lambda _{L}}$ (see (\ref{fock})), defined
by 
\begin{equation}
\tau _{t}^{(L)}(A)\doteq \mathrm{e}^{it\mathrm{H}_{L}}A\mathrm{e}^{-it%
\mathrm{H}_{L}}\ ,\qquad A\in \mathcal{B}(\mathcal{F}_{\Lambda _{L}}),\ t\in 
{\mathbb{R}}\ .  \label{dynamics full}
\end{equation}%
The generator of this time evolution is the linear operator $\delta _{L}$
defined on $\mathcal{B}(\mathcal{F}_{\Lambda _{L}})$ by%
\begin{equation*}
\delta _{L}\left( A\right) \doteq i[\mathrm{H}_{L},A]\doteq i\left( \mathrm{H%
}_{L}A-A\mathrm{H}_{L}\right) \ ,\qquad A\in \mathcal{B}(\mathcal{F}%
_{\Lambda _{L}})\ .
\end{equation*}%
If $\gamma =0$ then it is well-known that the thermodynamic limit of $\{\tau
_{t}^{(L)}\}_{t\in {\mathbb{R}}}$ exists as a strongly continuous group $%
\{\tau _{t}\}_{t\in {\mathbb{R}}}$ of $\ast $-automorphisms of the CAR
algebra of the infinite lattice. If $\gamma >0$ then the situation is not
that obvious. A first guess is to approximate $\{\tau _{t}^{(L)}\}_{t\in {%
\mathbb{R}}}$ by $\{\tau _{t}^{(L,c)}\}_{t\in {\mathbb{R}}}$, where 
\begin{equation}
\tau _{t}^{(L,c)}(A)\doteq \mathrm{e}^{it\mathrm{H}_{L}\left( c\right) }A%
\mathrm{e}^{-it\mathrm{H}_{L}\left( c\right) }\ ,\qquad A\in \mathcal{B}(%
\mathcal{F}_{\Lambda _{L}}),\ t\in {\mathbb{R}}\ ,  \label{dynamics approx}
\end{equation}%
for any $L\in \mathbb{N}_{0}$ and some complex number $c\in \mathbb{C}$. In
this case, the linear operator 
\begin{equation}
\delta _{L,c}\left( A\right) \doteq i[\mathrm{H}_{L}\left( c\right) ,A]\
,\qquad A\in \mathcal{B}(\mathcal{F}_{\Lambda _{L}})\ ,
\label{generator approx}
\end{equation}%
is the generator of the dynamics $\{\tau _{t}^{(L,c)}\}_{t\in {\mathbb{R}}}$%
. A natural choice for $c\in \mathbb{C}$ would be a solution to the
variational problem (\ref{var pb}), but what about if the solution is not
unique?\ As a matter of fact, as explained in \cite[Section 4.3]%
{BruPedra-MFII}, in the thermodynamic limit $L\rightarrow \infty $, the
finite-volume dynamics $\{\tau _{t}^{(L)}\}_{t\in {\mathbb{R}}}$ does \emph{%
not} converge within the CAR $C^{\ast }$-algebra of the infinite lattice for 
$\gamma >0$, even if $\mathfrak{d}=0$ would be the unique solution to the
variational problem (\ref{var pb})! Observe, moreover, that the variational
problem (\ref{var pb}) depends on the temperature whereas the time evolution
(\ref{dynamics full}) does not.

The validity of the Bogoliubov approximation (\ref{dynamics approx}) with
respect to the full dynamics (\ref{dynamics full}) was an open question that
Thirring and Wehrl \cite{T1,T2} solve in 1967 for the special case $\mathrm{H%
}_{L}|_{\mu =\lambda =h=0}$, which is an exactly solvable
permutation-invariant model for any $\gamma \in \mathbb{R}$. An attempt to
generalize Thirring and Wehrl's results to a general class of fermionic
models, including the BCS theory, has been done in 1978 \cite{Hemmen78}, but
at the cost of technical assumptions that are difficult to verify in
practice.\ This research direction has been strongly developed by many
authors until 1992, see \cite%
{Bona75,Sewell83,Rieckers84,Morchio87,Bona87,Duffner-Rieckers88,Bona88,Bona89,Bona90,Unnerstall90,Unnerstall90b,Unnerstall90-open,Bona91,Duffield1991,BagarelloMorchio92,Duffield-Werner1,Duffield-Werner2,Duffield-Werner3,Duffield-Werner4,Duffield-Werner5}%
. All these papers study dynamical properties of \emph{permutation-invariant}
quantum-spin systems with mean-field interactions. Our results \cite%
{BruPedra-MFII,BruPedra-MFIII} represent a significant generalization of
such previous results to possibly non-permutation-invariant lattice-fermion
or quantum-spin systems. To understand what's going on in the
infinite-volume dynamics, we now come back to our pedagogical example, that
is, the strong-coupling BCS-Hubbard model.

\section{Self-Consistency Equations}

Instead of considering the Heisenberg picture, let us consider now the Schr%
\"{o}dinger picture of quantum mechanics. In this case, recall that, at
fixed $L\in \mathbb{N}_{0}$, a finite-volume state $\rho ^{(L)}$ is defined
by 
\begin{equation*}
\rho ^{(L)}\left( A\right) \doteq \mathrm{Trace}_{\mathcal{F}_{\Lambda
_{L}}}\left( \mathrm{d}^{(L)}A\right) \ ,\qquad A\in \mathcal{B}(\mathcal{F}%
_{\Lambda _{L}})\ ,
\end{equation*}%
for a uniquely defined positive operator $\mathrm{d}^{(L)}\in \mathcal{B}(%
\mathcal{F}_{\Lambda _{L}})$ satisfying $\mathrm{Trace}_{\mathcal{F}%
_{\Lambda _{L}}}(\mathrm{d}^{(L)})=1$ and named the density matrix of $\rho
^{(L)}$. Compare with (\ref{gibbs1}) and (\ref{gibbs2}). At $L\in \mathbb{N}%
_{0}$, the time evolution of any finite-volume state is%
\begin{equation}
\rho _{t}^{(L)}\doteq \rho ^{(L)}\circ \tau _{t}^{(L)}\ ,\qquad t\in {%
\mathbb{R}}\ ,  \label{rho}
\end{equation}%
which corresponds to a time-dependent density matrix equal to $\mathrm{d}%
_{t}^{(L)}=\tau _{-t}^{(L)}(\mathrm{d}^{(L)})$.

The thermodynamic limit of (\ref{rho}) for periodic initial states can be
explicitly computed, as explained in \cite[Section 4.3.2]{Bru-pedra-MF-IV}.
It refers to a \emph{non-linear} state-dependent dynamics related to \emph{%
self-consistency}: By (\ref{fock}), $\mathcal{B}\left( \mathcal{F}%
_{\{0\}}\right) $ can be identified with the set $\mathrm{Mat}(4,\mathbb{C})$
of complex $4\times 4$ matrices, in some orthonormal basis\footnote{%
For instance, $\left( 1,0,0,0\right) $ is the vacuum; $\left( 0,1,0,0\right) 
$ and $\left( 0,0,1,0\right) $ correspond to one fermion with spin $\uparrow 
$ and $\downarrow $, respectively; $\left( 0,0,0,1\right) $ refers to two
fermions with opposite spins.}. For any continuous family $\omega \doteq
(\omega _{t})_{t\in \mathbb{R}}$ of states acting on $\mathcal{B}\left( 
\mathcal{F}_{\{0\}}\right) $, we define the finite-volume \emph{%
non-autonomous} dynamics $(\tau _{t,s}^{(L,\omega )})_{_{s,t\in \mathbb{R}}}$
by the Dyson-Phillips series 
\begin{equation*}
\tau _{t,s}^{(L,\omega )}\equiv \text{\textquotedblleft }\exp \left(
\int_{s}^{t}\delta _{L}^{\omega _{u}}\mathrm{d}u\right) \text{%
\textquotedblright }\doteq \mathbf{1}_{\mathcal{B}(\mathcal{F}_{\Lambda
_{L}})}+\sum\limits_{k\in {\mathbb{N}}}\int_{s}^{t}\mathrm{d}t_{1}\cdots
\int_{s}^{t_{k-1}}\mathrm{d}t_{k}\delta _{L}^{\omega _{t_{k}}}\circ \cdots
\circ \delta _{L}^{\omega _{t_{1}}}
\end{equation*}%
acting on $\mathcal{B}(\mathcal{F}_{\Lambda _{L}})$ for any $s,t\in \mathbb{R%
}$, with $\mathbf{1}_{\mathcal{B}(\mathcal{F}_{\Lambda _{L}})}$ being the
identity mapping of $\mathcal{B}(\mathcal{F}_{\Lambda _{L}})$ and where $%
\delta _{L}^{\rho }$ is the generator of the group $\{\tau
_{t}^{(L,c)}\}_{t\in {\mathbb{R}}}$, defined by (\ref{generator approx}) for 
$c=\rho (a_{0,\uparrow }a_{0,\downarrow })$. Compare with (\ref{dynamics
approx0}) and (\ref{dynamics approx00}). Note that, for every continuous
family $\omega \doteq (\omega _{t})_{t\in \mathbb{R}}$ of on-site (even)
states acting on $\mathcal{B}\left( \mathcal{F}_{\{0\}}\right) $, $s,t\in 
\mathbb{R}$, $L_{0}\in \mathbb{N}_{0}$ and all integers $L\geq L_{0}$, 
\begin{equation}
\tau _{t,s}^{(L,\omega )}\left( A\right) =\tau _{t,s}^{(L_{0},\omega
)}\left( A\right) \ ,\qquad A\in \mathcal{B}(\mathcal{F}_{\Lambda
_{L_{0}}})\ .  \label{idiot}
\end{equation}%
It follows that the family $\{\tau _{t,s}^{(L,\omega )}\}_{s,t\in {\mathbb{R}%
}}$ strongly converges in the thermodynamic limit $L\rightarrow \infty $ to
a strongly continuous two-parameter family $\{\tau _{t,s}^{\omega
}\}_{s,t\in {\mathbb{R}}}$ of $\ast $-auto%
\-%
morphisms of the CAR algebra\ of the lattice. With these observations, we
are in a position to give the self-consistency equations: By \cite[Eq. (19)]%
{Bru-pedra-MF-IV}, for any fixed initial (even) state $\rho $ on $\mathcal{B}%
\left( \mathcal{F}_{\{0\}}\right) $ at $t=0$, there is a unique family $(%
\mathbf{\varpi }(t,\rho ))_{t\in \mathbb{R}}$ of (on-site) states acting on $%
\mathcal{B}\left( \mathcal{F}_{\{0\}}\right) $ such that 
\begin{equation}
\mathbf{\varpi }(t,\rho )=\rho \circ \tau _{t,0}^{\mathbf{\varpi }(\cdot
,\rho )}\ ,\qquad t\in {\mathbb{R}}\ .  \label{self-consitency}
\end{equation}%
Observe that (\ref{self-consitency}) is an equation on a finite-dimensional
space, see (\ref{fock}).

\section{Infinite-Volume Dynamics of Product States}

For simplicity, as initial state (at $t=0$), take a finite-volume product%
\footnote{%
The product state $\rho ^{(L)}$ is (well-) defined by $\rho ^{(L)}(\alpha
_{x_{1}}(A_{1})\cdots \alpha _{x_{n}}(A_{n}))=\rho (A_{1})\cdots \rho (A_{n})
$ for all $A_{1},\ldots ,A_{n}\in \mathcal{B}\left( \mathcal{F}%
_{\{0\}}\right) $ and all $x_{1},\ldots ,x_{n}\in \Lambda _{L}$ such that $%
x_{i}\not=x_{j}$ for $i\not=j$, where $\alpha _{x_{j}}(A_{j})\in \mathcal{B}%
\left( \mathcal{F}_{\{x_{j}\}}\right) $ is the $x_{j}$-translated copy of $%
A_{j}$ for all $j\in \{1,\ldots ,n\}$.} state $\rho ^{(L)}\doteq \otimes
_{\Lambda _{L}}\rho $ associated with an even\footnote{%
Even states are the physically relevant ones. Even means that the
expectation value of any odd monomials in $\{a_{0,\mathrm{s}}^{\ast },a_{0,%
\mathrm{s}}\}_{\mathrm{s}\in \{\uparrow ,\downarrow \}}$ with respect to the
on-site state $\rho $ is zero.} state $\rho $ on $\mathcal{B}\left( \mathcal{%
F}_{\{0\}}\right) $. An example of finite-volume product states is given by
the approximating Gibbs states (\ref{gibbs2}). Then, in this case, as
explained in \cite[Section 4.4]{Bru-pedra-MF-IV}, for any $t\in \mathbb{R}$, 
$L_{0}\in \mathbb{N}_{0}$ and $A\in \mathcal{B}(\mathcal{F}_{\Lambda
_{L_{0}}})$, one has that 
\begin{equation}
\rho _{t}\left( A\right) \doteq \lim_{L\rightarrow \infty }\rho
_{t}^{(L)}\left( A\right) =\lim_{L\rightarrow \infty }\rho ^{(\infty )}\circ
\tau _{t}^{(L)}\left( A\right) =\rho ^{(\infty )}\circ \tau _{t,0}^{\mathbf{%
\varpi }(\cdot ,\rho )}\left( A\right) \ ,  \label{eq restrictedsimple}
\end{equation}%
with $\rho _{t}^{(L)},\mathbf{\varpi }(\cdot ,\rho )$ being respectively
defined by (\ref{rho}) and (\ref{self-consitency}) and where $\rho ^{(\infty
)}\doteq \otimes _{\mathbb{Z}^{d}}\rho $ is the (infinite-volume) product
state associated with the even state $\rho $ on $\mathcal{B}\left( \mathcal{F%
}_{\{0\}}\right) $.

For any $t\in \mathbb{R}$, the limit state $\rho _{t}$ is again a product
state and hence, it is completely determined by its restriction to the
single lattice site $(0,\ldots ,0)\in \mathbb{Z}^{d}$, that is, by the
on-site state $\mathbf{\varpi }(t,\rho )$ for all $t\in \mathbb{R}$. Below,
we give the explicit expressions for the time evolution of the most
important physical quantities related to this model, in this situation for
any time $t\in \mathbb{R}$:

\begin{itemize}
\item[(i)] Electron density:%
\begin{equation*}
\mathrm{d}(\rho )\doteq \rho \left( n_{0,\uparrow }+n_{0,\downarrow }\right)
=\rho _{t=0}\left( n_{0,\uparrow }+n_{0,\downarrow }\right) =\rho _{t}\left(
n_{0,\uparrow }+n_{0,\downarrow }\right) \in \lbrack 0,2].
\end{equation*}

\item[(ii)] Magnetization density: 
\begin{equation*}
\mathrm{m}(\rho )\doteq \rho \left( n_{0,\uparrow }-n_{0,\downarrow }\right)
=\rho _{t=0}\left( n_{0,\uparrow }-n_{0,\downarrow }\right) =\rho _{t}\left(
n_{0,\uparrow }-n_{0,\downarrow }\right) \in \lbrack -1,1].
\end{equation*}

\item[(iii)] Coulomb correlation density: 
\begin{equation*}
\mathrm{w}(\rho )\doteq \rho \left( n_{0,\uparrow }n_{0,\downarrow }\right)
=\rho _{t=0}\left( n_{0,\uparrow }n_{0,\downarrow }\right) =\rho _{t}\left(
n_{0,\uparrow }n_{0,\downarrow }\right) \in \lbrack 0,1].
\end{equation*}

\item[(iv)] Cooper field and condensate densities:%
\begin{equation*}
\rho _{t}\left( a_{0,\downarrow }a_{0,\uparrow }\right) =\sqrt{\mathrm{%
\kappa }(\rho )}\mathrm{e}^{i\left( t\mathrm{\nu }(\rho )+\theta (\rho
)\right) }\quad \text{with}\quad \mathrm{\nu }(\rho )\doteq 2\left( \mu
-\lambda \right) +\gamma \left( 1-\mathrm{d}(\rho )\right)
\end{equation*}%
and $\mathrm{\kappa }(\rho )\in \lbrack 0,1]$, $\theta (\rho )\in \lbrack
-\pi ,\pi )$ such that $\rho \left( a_{0,\downarrow }a_{0,\uparrow }\right) =%
\sqrt{\mathrm{\kappa }(\rho )}\mathrm{e}^{i\theta (\rho )}$.
\end{itemize}

\noindent See \cite[Lemma 1]{Bru-pedra-MF-IV}. In the special case $\lambda
=0$, i.e., without the Hubbard interaction, (i)-(iv) reproduce the results
of \cite[Section A]{Bona89} on the strong-coupling BCS\ model, written in
that paper as a permutation-invariant quantum-spin model.

From Assertions (i)-(iv) observe that we recover the equation of a symmetric 
\emph{rotor} in classical mechanics: Fix an even on-site state $\rho $. For
any $t\in \mathbb{R}$, define the 3D vector $(\Omega _{1}(t),\Omega
_{2}(t),\Omega _{3}(t))$ by 
\begin{equation*}
\rho _{t}\left( a_{0,\downarrow }a_{0,\uparrow }\right) =\Omega
_{1}(t)+i\Omega _{2}(t)\quad \text{and}\quad \Omega _{3}\left( t\right)
\doteq 2\left( \mu -\lambda \right) +\gamma \left( 1-\rho _{t}\left(
n_{0,\uparrow }+n_{0,\downarrow }\right) \right) .
\end{equation*}
Then, this 3D vector satisfies, for any time $t\in \mathbb{R}$, the
following system of ODEs:%
\begin{equation*}
\left\{ 
\begin{array}{l}
\dot{\Omega}_{1}\left( t\right) =-\Omega _{3}\left( t\right) \Omega
_{2}\left( t\right) \ , \\ 
\dot{\Omega}_{2}\left( t\right) =\Omega _{3}\left( t\right) \Omega
_{1}\left( t\right) \ , \\ 
\dot{\Omega}_{3}\left( t\right) =0\ ,%
\end{array}%
\right.
\end{equation*}%
which describes the time evolution of the angular momentum of a symmetric
rotor in classical mechanics.\ 

In fact, by seeing quantum states as elements of a state space in classical
mechanics, this dynamics can be written in terms of Poisson brackets, i.e.,
as some \emph{Liouville's equation} of classical mechanics, as proven in 
\cite[Corollary 6.11]{BruPedra-MFII} for any translation invariant
long-range models. Moreover, \cite{BruPedra-MFII,BruPedra-MFIII} show that
long-range dynamics in infinite volume are equivalent to intricate
combinations of classical and quantum short-range dynamics, opening new
theoretical perspectives, as explained in \cite{Bru-pedra-MF-I}. This
phenomenon is a direct consequence of the highly non-local character of
long-range, or mean-field, interactions.

Assertions (i)-(iv) lead to the exact dynamics of a physical system prepared
in a product state at initial time, driven by the strong-coupling
BCS-Hubbard Hamiltonian. This set of states is still [**restrictive**] and
our results \cite{BruPedra-MFII,BruPedra-MFIII} go beyond this simple case,
by allowing us to consider \emph{general periodic states as initial states},
in contrast with all previous results on lattice Fermi, or quantum-spin,
systems with long-range, or mean-field, interactions. See \cite[Section 2.6]%
{Bru-pedra-MF-IV}. \bigskip

\noindent \textit{Acknowledgments:} This work is supported by CNPq
(308337/2017-4), FAPESP (2017/22340-9), as well as by the Basque Government
through the grant IT641-13 and the BERC 2018-2021 program, and by the
Spanish Ministry of Science, Innovation and Universities: BCAM Severo Ochoa
accreditation SEV-2017-0718, MTM2017-82160-C2-2-P.

\end{document}